\documentclass[10pt]{iopart}

\usepackage{graphicx}
\usepackage{textcomp}
\usepackage[super,sort&compress,comma]{natbib}
\usepackage[utf8]{inputenc}
\usepackage{color}
\usepackage{pdfpages}
\usepackage{setstack}
\usepackage{gensymb}

\begin{document}
\title[Ultrafast charge-transfer dynamics in twisted MoS$_2$/WSe$_2$ heterostructures]{Ultrafast charge-transfer dynamics in twisted MoS$_2$/WSe$_2$ heterostructures}

\date{\today}

\author{J.~E. Zimmermann$^1$, M. Axt$^1$, F. Mooshammer$^2$, P. Nagler$^2$, C. Sch{\"u}ller$^2$, T. Korn$^3$,  U. Höfer$^1$ and G. Mette$^1$}

\address{$^1$ Fachbereich Physik und Zentrum f{\"u}r
Materialwissenschaften, Philipps-Universit{\"a}t, 35032 Marburg,
Germany}
\address{$^2$ Institut für Experimentelle und Angewandte Physik, Universit{\"a}t Regensburg, 93053 Regensburg, Germany}
\address{$^3$ Institut für Physik, Universit{\"a}t Rostock, 18059 Rostock, Germany}

\ead{gerson.mette@physik.uni-marburg.de}

\begin{abstract}
Two-dimensional transition metal dichalcogenides (TMD) offer a
unique platform for creating van-der-Waals heterojunctions with
fascinating physical properties and promising applications in
optoelectronics and valleytronics. Because of their typical
type-II band alignment, photoexcited electrons and holes can
separate via interfacial charge transfer. To understand the nature
and the dynamics of this charge transfer is of utmost importance
for the design and efficiency of potential devices. However,
systematic studies concerning the influence of the stacking angle
on the charge transfer remain sparse. Here, we apply time- and
polarization resolved second-harmonic imaging microscopy to
investigate the charge-transfer dynamics for three MoS$_2$/WSe$_2$
heterostructures with different stacking angles at a previously unattainable
time-resolution of $\approx$\,6\,fs. For 1.70\,eV excitation energy, electron transfer from WSe$_2$ to MoS$_2$ is found to
depend considerably on the stacking angle with the fastest
transfer time observed to be as short as 12\,fs. At 1.85\,eV excitation energy, ultrafast hole
transfer from MoS$_2$ to hybridized states at the $\Gamma$-point
or to the K-points of WSe$_2$ has to be considered. Surprisingly,
the corresponding decay dynamics show only a minor stacking-angle
dependence indicating that radiative recombination of indirect
$\Gamma$-K excitons becomes the dominant decay route for all
samples.
\end{abstract}


%\submitto{\TDM}

\maketitle

\ioptwocol

%%%%%%%%%%%%%%%%%%%%%%%%%%%%%%%%%%%%%%%%%%%%%%%%%%%%%%%%%%%%%%%
\section{Introduction}
%%%%%%%%%%%%%%%%%%%%%%%%%%%%%%%%%%%%%%%%%%%%%%%%%%%%%%%%%%%%%
Two-dimensional van-der-Waals materials offer a plethora of
opportunities for the design and investigation of stacked
heterostructures (HS)~\cite{Geim13nat,Lim14cm}. In particular,
heterostructures of transition metal dichalcogenides (TMD) have
revealed fascinating properties stimulating fundamental and applied
research in the field of optoelectronics and
valleytronics~\cite{Mak16natphot,Xu14natphys}. Many combinations of
different TMD layers form type-II heterojunctions~\cite{Oezce16prb}
which enables efficient charge separation and results in spatially
separated electron-hole pairs following an optical excitation (so
called charge-transfer or interlayer excitons)\
\cite{Wang21nl,Zheng20jpcl,Zimmer20,Calman20nl,Ovesen19,Kunstm18natphys,Rivera18natnano,Miller17nl,Schaib16natcomm,Rivera15natcomm,Fang14pnas,Hong14natnano}.
Due to their van-der-Waals coupled nature, structures with arbitrary stacking angles can be designed and manufactured. The relative
orientation of the TMDs in turn influences the coupling between the
layers and therefore charge transfer, recombination and other
properties of the interlayer
excitons~\cite{Brot20natmat,Merkl20natcomm,Gogoi19acsnano,Kunstm18natphys,Hsu18natcomm,Yeh16nl,Zheng15,Liu14natcomm}.
However, recent experimental studies investigating the influence of
the stacking on the ultrafast charge transfer obtained surprisingly
diverse results. For MoS$_2$/WSe$_2$ heterostructures, charge
transfer has been reported to be faster than the experimental time
resolution of 40\,fs independent of the examined stacking
angles~\cite{Zhu17nl}. In contrast, much slower charge transfer
times of a few hundred femtoseconds and a significant increase for
larger rotational mismatch were observed for
WS$_2$/WSe$_2$~\cite{Merkl19natmat}.

Thus, the underlying mechanism for the charge transfer process for different van-der-Waals heterostructures
still remains elusive and there are several
unresolved questions concerning the influence of the stacking
configuration on the ultrafast interlayer charge transfer. We have
recently shown the capabilities of the experimental method of time-
and polarization-resolved second-harmonic (SH) imaging microscopy
giving us direct access to the directional ultrafast charge transfer
in a rotationally mismatched WSe$_2$/MoSe$_2$
heterostructure~\cite{Zimmer20}. In particular, our technique allows
for pump-probe experiments in $\mu$m-sized regions of heterostructures with a
time-resolution not accessible in previous studies. Here, we employ this new approach to
examine the charge-carrier dynamics in MoS$_2$/WSe$_2$
heterostructures with different stacking angles. The three
MoS$_2$/WSe$_2$ samples studied in the present work had been
examined beforehand in a systematic study by
Kunstmann~\emph{et~al.}~\cite{Kunstm18natphys}. There it has been
conclusively shown that for all stacking configurations
photoluminescence from momentum-space indirect $\Gamma$-K interlayer
excitons is visible after optical excitation with 532\,nm (2.33\,eV)
photons. It is therefore established that for excitation
energies significantly exceeding the respective band gaps interlayer charge transfer occurs and that at least some of
the excited charge carriers recombine radiatively via a $\Gamma$-K
transition. Recently, also momentum-space direct K-K
photoluminescence at $\approx$1.0\,eV has been observed for
well-aligned MoS$_2$/WSe$_2$ heterostructures~\cite{Karni19prl}. In
the present study, we want to elucidate the corresponding ultrafast
dynamics in dependence of the stacking angle for three selected
stacking angles: 9\degree\ (referred to as quasi-3R), 52\degree\ (quasi-2H) and
16\degree\ (misaligned).

%%%%%%%%%%%%%%%%%%%%%%%%%%%%%%%%%%%%%%%%%%%%%%%%%%%%%%%%%%%%%%%
\section{Results and Discussion}%
%%%%%%%%%%%%%%%%%%%%%%%%%%%%%%%%%%%%%%%%%%%%%%%%%%%%%%%%%%%%%%%
%
\begin{figure*}[ht]%
\centering
\includegraphics{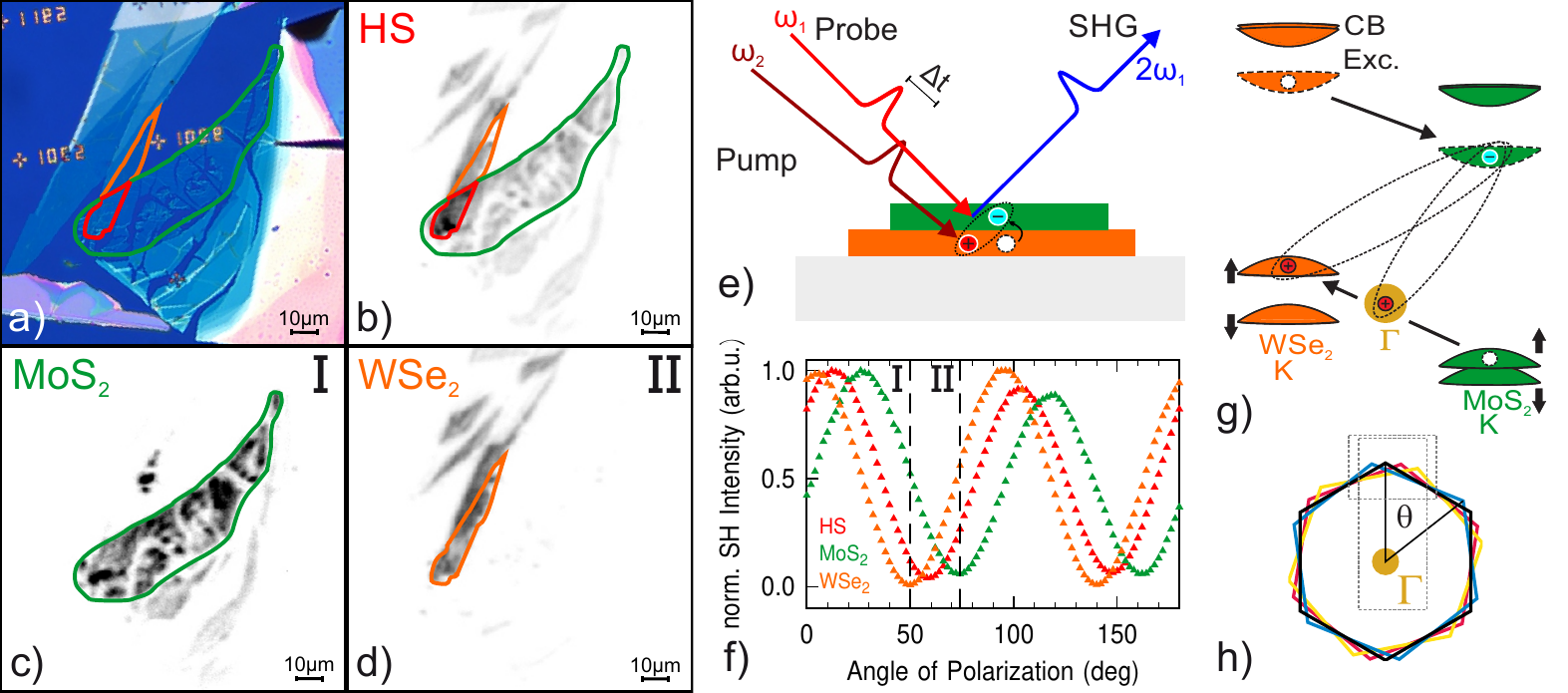}%[width=8.5cm]
%\captionsetup{font=scriptsize}
\caption{(a) Optical microscopy image of the studied misaligned (16\degree)
MoS$_2$/WSe$_2$ sample. MoS$_2$ and WSe$_2$ monolayer as well as the
heterostructure regions are highlighted with green, orange and red outlines, respectively.
(b)-(d) SH images measured with three different probe polarizations
sensitive to both monolayers (b), exclusively to MoS$_2$ (c) or WSe$_2$ (d), respectively.
(e)~Sketch of the pump-probe experiment for an example of resonant optical excitation in WSe$_2$
and selective SHG detection of potential charge transfer in MoS$_2$.
(f)  Polarisation-dependent SHG of the misaligned sample shown in (a)
evaluated for MoS$_2$ (green) and WSe$_2$ (orange) monolayers
and the heterostructure (red). The dashed lines mark the
polarization angles I and II applied for the SH images shown in (c) and (d).
(g) Schematic band alignment of MoS$_2$/WSe$_2$ showing the individual K-points of both materials
and the hybridized $\Gamma$-point~\cite{Kunstm18natphys}. The dashed
ellipses mark the previously reported K-K and $\Gamma$-K interlayer excitons~\cite{Kunstm18natphys,Karni19prl}. (h) Corresponding hexagonal Brillouin zones of the studied heterostructures for the three stacking angles 9\degree\ (red), 16\degree\ (yellow) and 52\degree\ (blue). The dashed rectangles delineate the regions of interest highlighted in Figs. 2(d) and 3(d) respectively.
\label{fig1}}
\end{figure*}%
Our method of choice is polarization- and time-resolved
second-harmonic imaging microscopy. By careful selection of the
polarization angle of the 800-nm (1.55\,eV) probe light, we are able to extract changes in the nonlinear susceptibility of individual monolayers
inside the TMD heterostructures. For TMD monolayers, the
pump-induced changes of the SH response upon resonant optical
excitation of intralayer excitons have been shown to be closely
correlated to the observed exciton dynamics in linear optical
spectroscopy~\cite{Zimmer20jp}. In case of TMD heterostructures, our
pump-probe experiments schematically sketched in Fig.~\ref{fig1}(e)
allow us to access the dynamics of interlayer charge
transfer~\cite{Zimmer20}. The three studied MoS$_2$/WSe$_2$ samples
were mechanically exfoliated and transferred onto a SiO$_2$/Si(001)
substrate as described in Ref.~\cite{Kunstm18natphys}.
Figs.~\ref{fig1}(a)-(d) compare an optical microscopy image of one
of the studied samples with respective images obtained for different
probe polarizations with our SH imaging microscopy. Due to the
inherent structural sensitivity of second-harmonic generation (SHG),
we are able to directly identify the different monolayer regions and
their overlap to determine the position of the heterostructure.

\begin{figure}[ht]
\centering
\includegraphics[width=7.9cm]{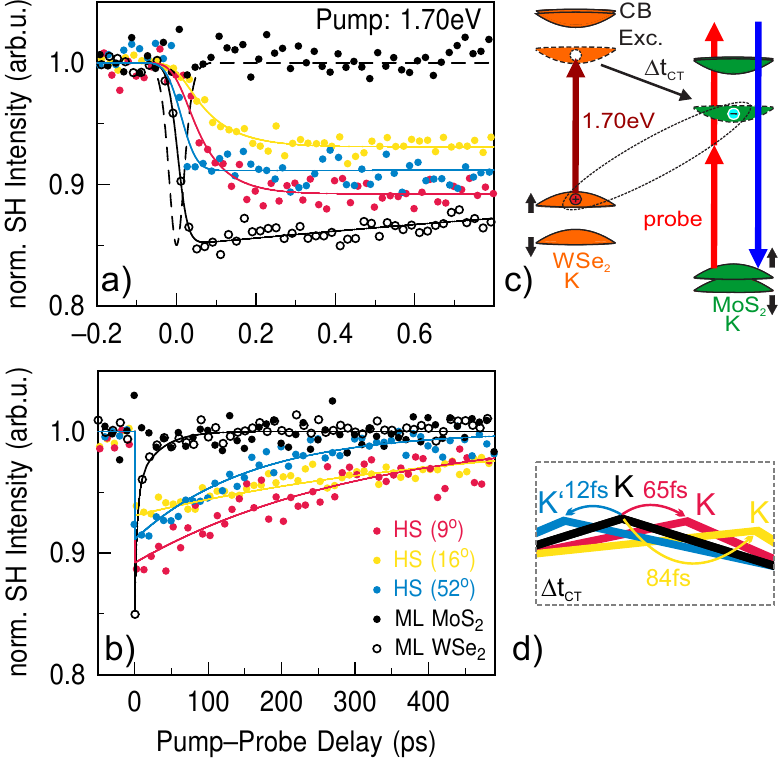}%[width=8.5cm]
%\captionsetup{font=scriptsize}
\caption{Time-resolved SHG of the MoS$_2$/WSe$_2$
heterostructures with three different stackings. (a,b) SH transients
upon 1.70\,eV resonant excitation of the
WSe$_2$ intralayer A-exciton for a sub-picosecond and an extended range of pump-probe delays, respectively.
Colored data points represent the heterostructure samples, black data points the corresponding monolayer dynamics as denoted in (b).
Solid lines in (a) correspond to rate-equation fits, those in (b) to exponential fits.
Filled (unfilled) data points are measured at probe polarization I (II),
sensitive to MoS$_2$ (WSe$_2$). The dashed line in (a) represents the cross-correlation (CCR) of the
laser pulses. (c) Sketch of the band alignment for MoS$_2$/WSe$_2$.
Ultrafast charge transfer to MoS$_2$ after resonant optical excitation
of WSe$_2$ results in the formation of K-K interlayer
excitons (dashed ellipse). SH detection selectively probes the
transient response of MoS$_2$. (d) Cutout from the Brillouin zone illustrating the electron transfer for the respective samples.
\label{fig2}}
\end{figure}

Fig.~\ref{fig1}(f) displays the normalized SH response in dependence
of the polarization of the probe laser evaluated for the monolayer
and heterostructure regions. From these measurements the crystal
orientations in relation to the lab coordinates can be determined.
The probe polarization for our time-resolved measurements is then
chosen to maximize the sensitivity for the MoS$_2$ layer by fully
suppressing the SH contribution of WSe$_2$ (polarization I). For
1.70\,eV pump-photon energy intralayer A-excitons of WSe$_2$ are
resonantly excited~\cite{Kozawa14natcomm,Li14prb}.
Fig.~\ref{fig2}(a) shows the corresponding SH-transients of the
heterostructures selectively detected by probing the ultrafast nonlinear response of the MoS$_2$ layer in
direct comparison with the respective transients obtained from the
individual monolayers. All SH transients have been normalized to the
signal at negative delays. As expected the MoS$_2$ monolayer
transient (filled black data) shows no pump-induced change since the
excitation energy is below its A-exciton resonance. In contrast, the SH-intensity of the MoS$_2$ layers of the heterostructures exhibits a clear modulation as a function of the pump-probe delay time. We observe a rapid decrease and a subsequent slower recovery. Since any direct excitation of
the MoS$_2$ can be excluded from the monolayer results, we can
therefore assign the observed dynamics in the heterostructures to ultrafast electron transfer to MoS$_2$ after optical excitation of WSe$_2$ as sketched
in Fig.~\ref{fig2}(c). This attribution is also corroborated by the
delayed onset of the heterostructure signals in relation to the WSe$_2$ monolayer response and by the ongoing/delayed decrease of the SH intensity, which extends even beyond the duration of the pump pulse (cf. dashed line, CCR).

While there is clear evidence for charge transfer in all
heterostructure samples, the observed ultrafast dynamics differ
considerably for the three different stackings. Apparently, the
fastest electron transfer can be found for the quasi-2H stacked
sample (blue data), followed by the quasi-3R (red data) and the
misaligned sample (yellow data). This is confirmed quantitatively
by a rate-equation model consisting of coupled differential
equations in which one energetically higher lying state is filled by
a gaussian pump pulse and then subsequently populates the lower
lying state with a transfer time $\Delta\mathrm{t}_{\mathrm{CT}}$. The extracted
transfer times for the quasi-3R and the misaligned sample are
comparable with values of 65$\pm$10\,fs and 84$\pm$10\,fs, respectively,
whereas the transfer time for the quasi-2H stacked sample is more
than five times faster (12$\pm$6\,fs). This disparity might be
surprising since the variation of the lattice separation between the
layers, which has a central influence on the tunneling
probability~\cite{Zhou20acsnano}, is negligible between 2H- and
3R-stacking~\cite{Kunstm18natphys}. However, it has also been
reported that the wavefunction overlap between the layers plays an
important role for the charge transfer across the interface as well
as for the subsequent recombination dynamics~\cite{Merkl20natcomm}.
Accordingly, wavefunction calculations for the interfacial plane in
TMD heterostructures have shown strong differences for 3R- and
2H-stacking~\cite{Wang16natcomm}, which then in turn can indeed
influence the dynamics~\cite{Zheng21pccp}.

Consequently, we conclude that the faster interlayer electron
transfer for the quasi-2H stacked sample is caused by the stronger interlayer
interaction of the involved monolayer states at the K-point. In accordance, the observed recombination dynamics shown in
Fig.~\ref{fig2}(b) also differ significantly for the three
heterostructures as determined from a single-exponential fit of the
SH transients to extract the individual lifetimes. The misaligned
sample shows the longest (467$\pm$16\,ps), the quasi-3R stacked
sample an intermediate (309$\pm$12\,ps) and the quasi-2H stacked
sample the shortest lifetime (158$\pm$12\,ps). The enhanced lifetime for the misaligned sample could be an indication for the absence of the radiative decay channel as was reported recently\cite{Karni19prl}. The difference in lifetime between quasi-2H and quasi-3R can then
be explained by the same reasoning as the enhanced transfer. Since
the charge transfer via a phonon-assisted tunneling process can be
amplified by the spatial overlap of the involved
wavefunctions~\cite{Wang21nl,Zhou20acsnano}, the same should hold
for the reversed process of interlayer exciton recombination: A
stronger spatial coincidence facilitates the recombination resulting
then in a reduced lifetime. Please note, that all three
heterostructure lifetimes are considerably extended in comparison to
the lifetime observed for the WSe$_2$ monolayer. The latter can be
described by a bi-exponential recovery behaviour with much shorter
lifetimes of $\tau_1$\,=\,3.5\,ps and $\tau_2$\,=\,30\,ps. This
enhanced lifetime for the heterostructure signals is in perfect
agreement with our interpretation, since the spatially indirect
nature of interlayer excitons leads to a strong increase of their
lifetimes in comparison to intralayer
excitons~\cite{Ceballos14acsnano, Rivera15natcomm}.

\begin{figure}[ht]
\centering
\includegraphics[width=7.9cm]{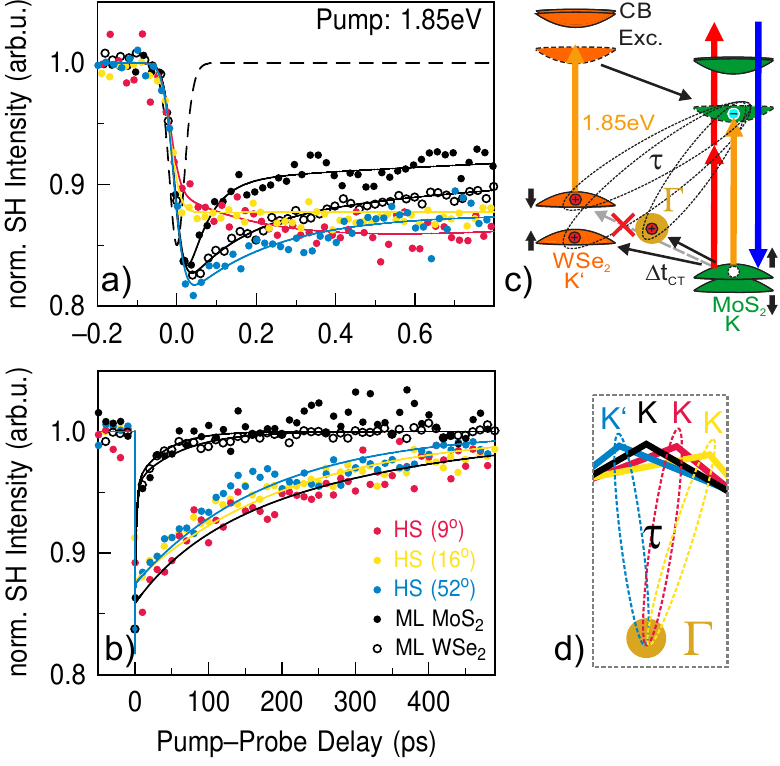}%
%\captionsetup{font=scriptsize}
\caption{Time-resolved SHG of the MoS$_2$/WSe$_2$ heterostructures with three
different stackings. (a,b) SH transients and corresponding fits as in Fig.~\ref{fig2}(a,b)
but for 1.85\,eV pump-photon energy, resonant with the MoS$_2$ intralayer A-exciton.
(c) Sketch of the energy alignment for MoS$_2$/WSe$_2$
following the same scheme as in Fig~\ref{fig2}(c), but
for resonant excitation in MoS$_2$ while still probing MoS$_2$. The
higher excitation energy opens up additional relaxation pathways by
ultrafast hole transfer to the hybridized
$\Gamma$-point or to the K/K'-points of WSe$_2$. Three different
kinds of conceivable interlayer excitons are marked by dashed ellipses.
For the quasi-2H stacked sample, hole transfer to K' is spin-forbidden and
only two transfer possibilities remain. (d) Cutout from the Brillouin zone illustrating the charge recombination via the $\Gamma$-point.
\label{fig3}}
\end{figure}

In the following, we would like to address open questions concerning
the ultrafast hole transfer and the particular influence that
hybridized states at the $\Gamma$-point have on charge transfer and
recombination. In order to elucidate these processes, the
pump-photon energy is tuned to 1.85\,eV for resonant optical
excitation of MoS$_2$ intralayer
A-excitons~\cite{Kozawa14natcomm,Li14prb}. While at this photon
energy both materials are excited, the generated exciton density
in MoS$_2$ (5.3$\cdot10^{12}/$cm$^2$) is about five times larger
than in WSe$_2$ (9.3$\cdot10^{11}/$cm$^2$), as calculated based on energy-dependent absorption of the two layers (see method section). Thus, the pump-induced
effects in WSe$_2$ are not negligible, however, the dynamic response
measured in MoS$_2$ is mainly dominated by its inherently generated
charge carriers. Fig.~\ref{fig3} shows the same time regimes as before using identical colors for the different structures, but now the systems are pumped at 1.85\,eV.
By comparing the dynamics of Figs.~\ref{fig2}(a) and~\ref{fig3}(a)
striking differences become obvious: First of all, upon 1.85-eV
photoexcitation the pump-induced effects now already start to occur
precisely at temporal overlap in accordance with a direct excitation
in MoS$_2$. With respect to the delayed filling, the SH transients
of the quasi-3R and the misaligned sample stay similar, whereas the
behaviour of the quasi-2H stacked sample changes drastically. In
contrast to the 1.70-eV excitation, where the SH signal of the
quasi-2H sample was observed to recover mono-exponentially with a
slow recovery rate, the SH response after 1.85-eV excitation begins
to recover immediately on a femtosecond timescale. The extracted
lifetime $(\tau_{\mathrm{2H}}\,=\,220\mathrm{\,fs})$ is only
slightly slower than the corresponding monolayer decay times for
1.85-eV excitation ($\tau_{\mathrm{MoS}_2}\,=\,75\mathrm{\,fs},\
\tau_{\mathrm{WSe}_2}\,=\,146\mathrm{\,fs}$).

This ultrafast decay process is associated with the coherent
radiative recombination of intralayer excitons~\cite{Selig18tdm, Poellm15natmat}. On
the one hand, the slightly slower recovery for the heterostructure
signal compared to the monolayer can be explained by the change of
the dielectric environment: The enhanced screening for the
heterostructure decreases the binding energy of the excitons and
therefore increases their Bohr radius and the recombination time. On
the other hand, the coherent radiative recombination competes with
additional relaxation mechanism in the heterostructure such as hole
transfer to the hybridized $\Gamma$-point or to the WSe$_2$ layer as
illustrated in Fig.~\ref{fig3}(c). Since at 1.85-eV excitation
electron transfer from the WSe$_2$ into the MoS$_2$ is feasible as well, the
detected SH signal in the MoS$_2$ is a mixture of various processes
and a clear interpretation of our SH transients becomes more
challenging. The competing processes might also explain why we do
not observe any clear ultrafast component for the quasi-3R and the
misaligned heterostructure sample in our data. Another plausible
explanation for the absence of the ultrafast decay in these two
heterostructures could also be related to the energy alignment and
the spin structure of the involved states. After photoexcitation
inside the MoS$_2$ the generated hole has two obvious relaxation
pathways alternative to intralayer recombination: it can either be
scattered to the hybridized states around the $\Gamma$-point or to
the K/K'-point in WSe$_2$. In the case of 2H-stacking, transfer to
the valence band maximum of WSe$_2$ at K' is spin forbidden as
sketched in Fig.~\ref{fig3}(c). For the misaligned sample,
scattering to K/K' should also be reduced due to the large momentum
mismatch. In contrast, hole transfer to the valence band maximum at
the K-point should be most efficient in case of 3R-stacking.
Assuming the hole transfer to K/K' to be a competing process to the
coherent recombination and taking place on a similar or faster
timescale, the decreasing efficiency of the hole transfer to K/K'
from quasi-3R, misaligned to quasi-2H could then explain the
increasing manifestation of the coherent recombination.

%Since electron transfer from WSe$_2$ to MoS$_2$ is much faster in the quasi-2H sample,
%Assuming the transfer to the $\Gamma$-point to be equally efficient
%for all heterostructure independent of the stacking angle, thus

The comparison of the SH transients at large pump-probe delays for
both excitation energies in Figs.~\ref{fig2}(b) and~\ref{fig3}(b)
reveals a surprising change in the dynamics caused by the higher
excitation energy. For 1.70-eV excitation the recovery rates of the
three heterostructure samples differ considerably as discussed
above. In the case of 1.85-eV excitation, however, the lifetimes
become very similar for all samples (quasi-2H:~178$\pm$9\,ps,
quasi-3R:~217$\pm$8\,ps, misaligned:~219$\pm$7\,ps). For 1.70-eV
excitation, the quasi-2H configuration showed the fastest recovery
time due to the larger wavefunction overlap between the layers. For
1.85\,eV, however, the lack of interaction between the layers for
the other two samples is compensated by the additional decay routes
of the charge carriers generated in the MoS$_2$. The central
difference between the two excitation energies is that for 1.85-eV
excitation the hole states around the $\Gamma$-point become energetically
available. Photoluminescence from momentum-space indirect $\Gamma$-K
interlayer excitons has been observed for all examined stacking
configurations of MoS$_2$/WSe$_2$~\cite{Kunstm18natphys}. Therefore,
we attribute our observed decay dynamics to the radiative
recombination of $\Gamma$-K interlayer excitons. Due to the fact
that the states around the $\Gamma$-point consist inherently of
orbitals which are delocalized and therefore spread out over both
layers the overlap of the wavefunctions of electron and hole
is enhanced and, thus, radiative recombination is facilitated. This
overall enhancement of the recombination compensates the reduced
overlap for the quasi-3R and misaligned sample equalizing the
lifetimes of the excitation independent of the stacking
configuration.

%%%%%%%%%%%%%%%%%%%%%%%%%%%%%%%%%%%%%%%%%%%%%%%%%%%%%%%%%%%%%
\section{Conclusion}
%%%%%%%%%%%%%%%%%%%%%%%%%%%%%%%%%%%%%%%%%%%%%%%%%%%%%%%%%%%%%
%
In conclusion, we have employed time- and polarization-resolved
second-harmonic imaging microscopy to study the ultrafast
charge-carrier dynamics across the MoS$_2$/WSe$_2$ heterostructure
interface for different stacking configurations. For lower
excitation energies of 1.70\,eV, electron transfer from WSe$_2$ to
MoS$_2$ is found to depend considerably on the stacking angle and
the transfer time is reduced by a factor of seven when going from a
larger rotational mismatch towards 2H-stacking. At higher
excitation energies, ultrafast hole transfer from MoS$_2$ to
hybridized states at the $\Gamma$-point and to the K-points of
WSe$_2$ has to be considered. The respective decay dynamics,
however, does not show a significant dependence on the stacking
angle indicating that radiative recombination of indirect $\Gamma$-K
excitons becomes the dominant decay route for all samples.
%
%%%%%%%%%%%%%%%%%%%%%%%%%%%%%%%%%%%%%%%%%%%%%%%%%%%%%%%%%%%%%
%%%%%%%%%%%%%%%%%%%%%%%%%%%%%%%%%%%%%%%%%%%%%%%%%%%%%%%%%%%%%%%
\ack Funding was provided by the Deutsche Forschungsgemeinschaft
(DFG, German Research Foundation), Project-ID 223848855-SFB 1083.
T.~Korn gratefully acknowledges funding by the DFG via
\mbox{KO3612/4-1}. C.~Schüller gratefully acknowledges funding by
the DFG via Project-ID 314695032-SFB 1277 as well as SPP2244
\mbox{(SCHU1171/10-1)}.
%%%%%%%%%%%%%%%%%%%%%%%%%%%%%%%%%%%%%%%%%%%%%%%%%%%%%%%%%%%%%%%

%%%%%%%%%%%%%%%%%%%%%%%%%%%%%%%%%%%%%%%%%%%%%%%%%%%%%%%%%%%%%%%
\section{Methods}
%%%%%%%%%%%%%%%%%%%%%%%%%%%%%%%%%%%%%%%%%%%%%%%%%%%%%%%%%%%%%%%%%%%%
%%%%%%%%%%%%%%%%%%%%%%%%%%%%%%%%%%%%%%%%%%%%%%%%%%%%%%%%

The experiments were performed under ambient conditions using a
Yb:KGW-based ultrashort pulse laser system (Light Conversion
Carbide) feeding two optical parametric amplifiers (Orpheus-F Twin,
Orpheus-N) providing photon energies in the range from 1.3-1.9\,eV
at pulse lengths routinely shorter than 35\,fs at a repetition rate
of 200\,kHz. The photon energy of the probe pulse was chosen to be
1.55\,eV well below the bandgap of both materials in order to
exclude optical excitation by the probe beam. Our temporal
resolution is only limited by the laser system and can be estimated
by measurement of the cross-correlation on the sample. The CCR
is measured to be 30\,fs. The minimal time-resolution is estimated
to be 1/5th of the CCR and corresponds to 6\,fs. Both beams are
focussed collinearly onto the sample under an angle of 18\degree.
After passing a 400-nm dielectric filter, the specular reflected SH
response of the probe beam is imaged optically magnified by a camera
lens on an electron-multiplied CCD chip (Princeton Instruments
ProEM-HS). Typical exposure times are 10\,seconds. The applied
magnification was M$\approx$35-40. The overall resolution of our
imaging microscopy setup is better than 4\,$\mu$m. The time-delay
between pump and probe beam is varied by a motorized delay stage.
The polarization of pump and probe beam can be varied by means of
$\lambda$/2-plates. The typically detected p-polarization of the
second-harmonic light is separated by an analyzer. The applied pump
fluence was fixed to 50\,$\mu$J/cm$^2$ for 1.85\,eV  and
38\,$\mu$J/cm$^2$ for 1.70\,eV on the sample (spot diameter 1/e).
The applied probe fluence was fixed to 250\,$\mu$J/cm$^2$. These
fluences lead to an initial exciton density of
3.65$\cdot$10$^{12}$/cm$^2$ (WSe$_2$) for 1.70\,eV and
9.3$\cdot$10$^{11}$/cm$^2$ (WSe$_2$) as well as
5.3$\cdot$10$^{12}$/cm$^2$ (MoS$_2$) for 1.85\,eV pump energy
(calculated with data from~\cite{Li14prb} with the tmm
package~\cite{tmm_package}). These excitation densities are below
the calculated Mott-density~\cite{Meckbach18apl} ensuring that we
are measuring exciton and not plasma dynamics. Long term
measurements with these fluences applied did not exhibit any
multishot damage. A more detailed description of the setup can be
found elsewhere~\cite{Zimmer20jp}.
Sample preparation and characterization is described in detail in Ref. \cite{Kunstm18natphys}.

%%%%%%%%%%%%%%%%%%%%%%%%%%%%%%%%%%%%%%%%%%%%%%%%%%%%%%%%%%%%%%%
%
%
%\clearpage
%
\newcommand{\newblock}{}
\bibliographystyle{unsrt}

\end{document}